\documentclass[slac_one]{revtex4}
\usepackage{graphicx}
\usepackage{fancyhdr}
\begin{document}

\title{Beyond the Standard Model Higgs at LHC} 

%

\author{Steven Lowette (on behalf of the ATLAS, CMS and FP420 Collaborations)}
\affiliation{University of California, Santa Barbara, CA 93106, USA}

\begin{abstract}
Models of Beyond the Standard Model (BSM) physics, like the Minimal Supersymmetric Standard Model (MSSM), often
involve an extended Higgs sector, giving rise to extra neutral or charged
Higgs bosons. The discovery reach expected from simulation studies for such
additional Higgs particles is presented for the ATLAS, CMS and FP420
detectors at the LHC. Emphasis is put on production and decay modes
involving heavy flavour b and tau particles, which are enhanced in large
regions of BSM parameter space. The LHC experiments are indeed particularly
well equipped to tackle final states containing heavy flavour.
\end{abstract}

\maketitle

\thispagestyle{fancy}

\section{INTRODUCTION}

In the Standard Model (SM), the Higgs mechanism and the still unobserved scalar Higgs boson form the cornerstone of electroweak symmetry breaking. Several problems of the SM, however, like the hierarchy problem, have triggered extensions to the SM with new symmetries, dimensions or interactions. At the Large Hadron Collider (LHC)~\cite{Evans:2008zz}, the new $14 \, \rm TeV$ ${\rm pp}$ collider at CERN, the general-purpose ATLAS~\cite{:2008zzm} and CMS~\cite{:2008zzk} experiments have been particularly well equipped to probe for the existence of Higgs bosons. In this report highlights are given of the beyond the Standard Model (BSM) Higgs boson search program at these detectors. Many of the search channels profit from the detectors' emphasis on heavy flavour ${\rm b}$ and $\tau$ identification. Also the possibilities with the FP420 detectors~\cite{Albrow:2008pn} are discussed, a proposed project to provide proton tracking and precise timing at $420 \, {\rm m}$ from the ATLAS and CMS interaction points, giving access to new particles in central exclusive production in the mass range $60 - 180 \, {\rm GeV}/c^2$.

\section{HIGGS BOSONS IN THE $CP$-CONSERVING MSSM}

The Minimal Supersymmetric Standard Model (MSSM) contains two Higgs doublets, one of which couples to up-type and one to down-type fermions. After electroweak symmetry breaking five physical particles remain: the neutral ${\rm h}$, ${\rm H}$ ($CP$-even) and ${\rm A}$ ($CP$-odd), and the charged ${\rm H}^{\pm}$ Higgs bosons. At the lowest order all Higgs couplings are controlled by two parameters, usually taken as the mass $m_{\rm A}$ and the ratio of the vacuum expectation values of the Higgs fields $\tan \beta$. At large $m_{\rm A}$ the lightest Higgs boson ${\rm h}$ becomes decoupled and SM-like. At large $\tan\beta$ the couplings to the up-type fermions ${\rm b}$ and ${\tau}$ are enhanced. Higher order corrections depend on the MSSM parameters, and are usually studied in particular benchmark scenarios, e.g. the $m_{\rm h}^{\rm max}$ scenario~\cite{Carena:2002qg}.


\subsubsection*{Neutral MSSM Higgs-Boson Searches}

Neutral MSSM Higgs-boson production happens dominantly through gluon fusion at small and moderate $\tan\beta$. ${\rm b\bar{b}}$-associated production dominates at large $\tan\beta$, and becomes particularly important at high Higgs-boson masses. The vector boson fusion (VBF) process becomes relevant with $m_{\rm h}$ at its maximum or $m_{\rm H}$ at its minimal mass limit. VBF production is not allowed for the pseudoscalar ${\rm A}$. The Higgs-strahlung processes, finally, with a ${\rm W}$ or ${\rm Z}$ boson produced in association, have smaller cross sections and suffer from large backgrounds at the LHC. 
Both the CMS and ATLAS collaborations have performed detailed studies of a plethora of search channels, of which some highlights are summarized in the following.

The branching fraction $\Phi \to \tau^{+}\tau^{-}$ ($\Phi = {\rm h,H,A}$) is of the order of 10\% over a wide parameter range, subleading to the very difficult $\Phi \to {\rm b\bar{b}}$. Search strategies for this tauonic decay depend on the further decay of the $\tau$ leptons, either hadronically or into electron or muon. In the associated production, ${\rm b\bar{b}}\Phi$ ($\Phi = {\rm h,H,A}$), with both taus decaying hadronically, $\Phi\to\tau^{+}\tau^{-}\to 2 \,\tau$-jets, selection of the signal at the trigger level becomes crucial. Also offline $\tau$-jet identification is important to reject jets from the multi-jet background~\cite{Gennai:2006bf}. This can be obtained with tracker isolation, and using the requirements of one or three tracks and a hard leading track. The QCD background shape can be obtained from data, using a sample of signal-free same-sign $\tau$ jets.

The search in the channel ${\rm b\bar{b}}\Phi$ ($\Phi = {\rm h,H,A}$), $\Phi\to\tau^{+}\tau^{-}\to {\rm e}/\mu + \tau$-jet can exploit the presence of an isolated electron or muon to trigger on, possibly combined with a $\tau$-jet. The main background comes from top production. In the CMS analyses~\cite{Kinnunen:2006ba,Kalinowski} this background is further reduced requiring only one $\rm b$-jet and vetoing on additional jets.

For the channel ${\rm b\bar{b}}\Phi$ ($\Phi = {\rm h,H,A}$), $\Phi\to\tau^{+}\tau^{-}\to 2 \ell$, ($\ell = {\rm e,\mu}$) ATLAS has explored a search including all electron and muon final state combinations~\cite{ATLASPerformance}. A dilepton mass window and a cut on missing transverse energy (MET) is applied to veto ${\rm Z \to e^{+}e^{-}/\mu^{+}\mu^{-}}$. The use of ${\rm b}$-tagging further suppresses ${\rm Z \to\tau^{+}\tau^{-}}$, while a soft leading jet requirement rejects top backgrounds. It is then possible to reconstruct the Higgs-boson mass, using the approximation of the neutrinos being collinear with the leptons. The remaining ${\rm Z\to\tau^{+}\tau^{-}}$ background is estimated from the data, while other systematics come from the jet energy scale and the knowledge of the ${\rm b}$-tagging efficiencies.

A different ${\rm b\bar{b}}\Phi$ ($\Phi = {\rm h,H,A}$) search channel looks for $\Phi\to\mu^{+}\mu^{-}$ decays. Although the branching fraction for this channel is very low in the Standard Model, it gets enhanced in the MSSM at large $\tan\beta$ (order $10^{-4}$). In the ATLAS study~\cite{ATLASPerformance} the Drell-Yan background can be suppressed with ${\rm b}$ tagging, while top pairs are rejected asking for small MET and soft jets. The good muon momentum resolution makes the Higgs boson mass peak stand out on a falling background, such that this background can be estimated from side bands. Also the Higgs boson width is measurable in this channel, which gives access to $\tan\beta$.

\begin{figure}[t]
  \begin{center}
    \begin{minipage}[t]{.48\textwidth}
      \begin{center}
        \includegraphics[scale=0.32]{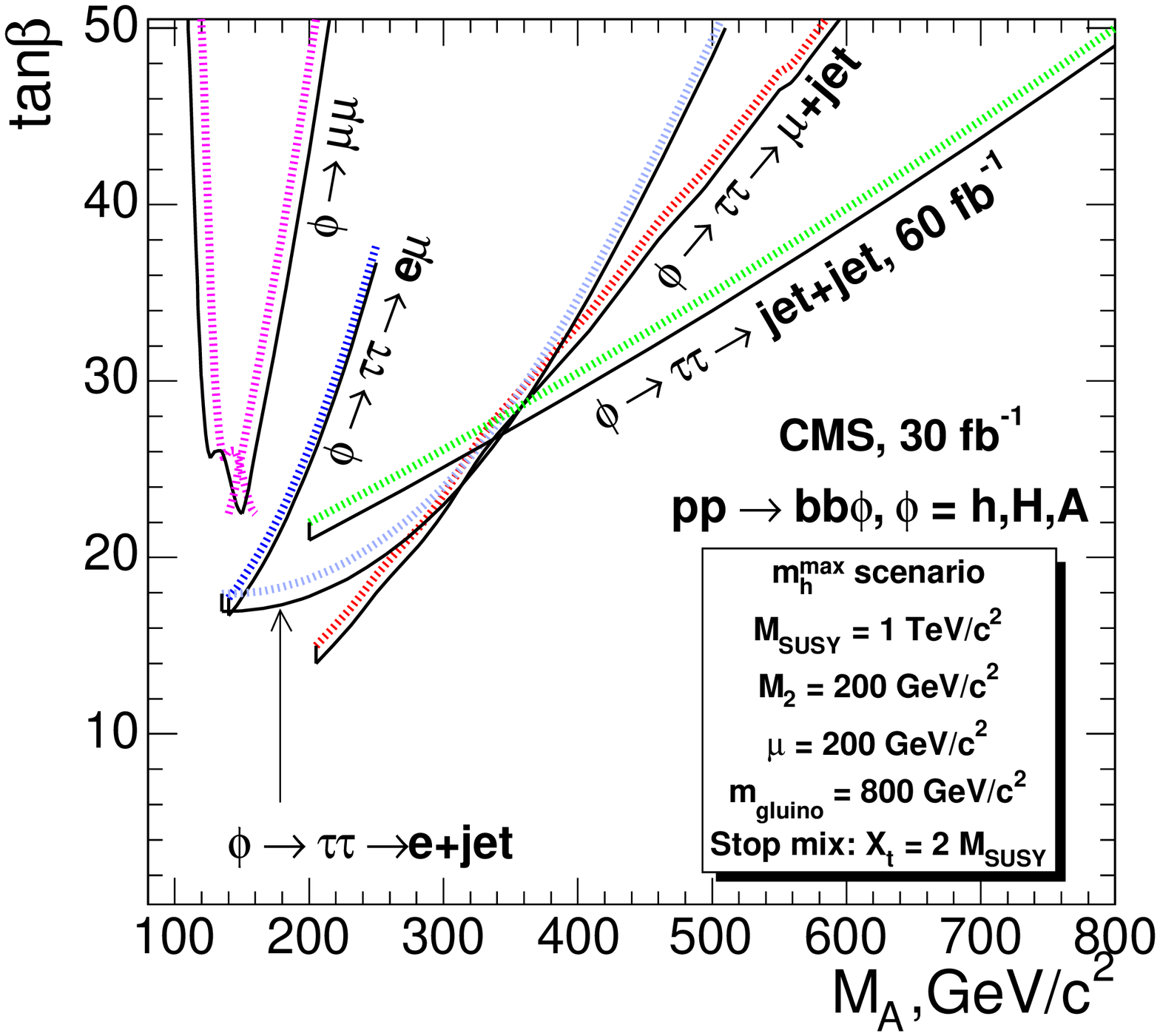}
        \vspace{-3mm}
        \caption{CMS discovery reach for MSSM associated ${\rm bb \Phi}$ production (${\Phi = h,H,A}$)~\cite{Ball:2007zza}.}
        \label{fig:cmsassoccontours}
      \end{center}
    \end{minipage}
    \hspace{\stretch{1}}
    \begin{minipage}[t]{.48\textwidth}
      \begin{center}
        \includegraphics[scale=0.32]{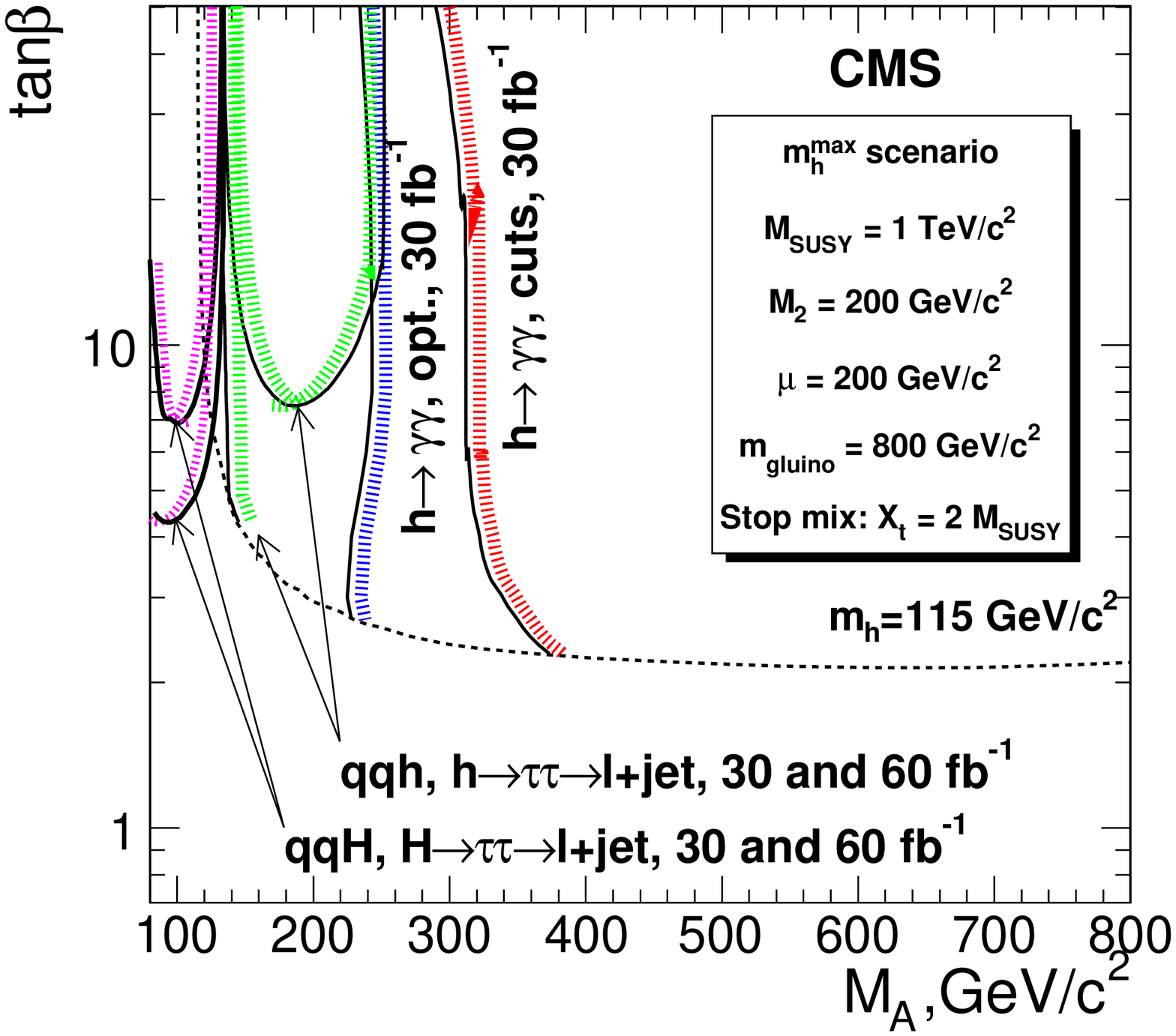}
        \vspace{-3mm}
        \caption{CMS discovery reach for MSSM ${\rm h\to\gamma\gamma}$ and ${\rm qqh/H}$, ${\rm h/H \to\tau\tau}$~\cite{Ball:2007zza}.}
        \label{fig:cmsvbfinclcontours}
      \end{center}
    \end{minipage}
  \end{center}
  \vspace{-5mm}
\end{figure}

In Figure~\ref{fig:cmsassoccontours} the discovery reach for all neutral Higgs boson associated production channels is summarized for the CMS experiment. In addition two searches for SM Higgs searches can be directly interpreted in the MSSM. For low $m_{\rm h}$ the inclusive ${\rm h\to\gamma\gamma}$ search~\cite{Bettinelli,Pieri:2006bm} uses the excellent calorimetry to resolve the di-photon mass peak on a large QCD background. The VBF $\rm q\bar{q}\Phi$ ($\Phi = {\rm h,H}$), $\Phi \to \tau^{+}\tau^{-}$ search~\cite{Foudas,Asai:2004ws}, on the other hand, is based on the identification of forward jets and a central jet veto. The discovery reach for these channels is shown in Figure~\ref{fig:cmsvbfinclcontours}.

Many other search strategies are pursued for neutral Higgs bosons, some with limited applicability or larger model dependence, but often looking for interesting signatures. The search for the ${\rm A \to Zh}$ decay, with ${\rm Z\to\ell\ell, h\to b\bar{b}}$~\cite{Anagnostou:2006du}, exploits the large ${\rm A\to Zh}$ branching fraction at low $\tan\beta$ and moderate $m_{\rm A}$, complemented with the dominant ${\rm h\to b\bar{b}}$ decay. Also decays into sparticles can be considered, for instance in the ${\rm A/H \to \chi_2^0\chi_2^0 \to 4 \ell + MET}$ decay~\cite{Charlot:2006se}. This is promising to cover the difficult low and intermediate $\tan\beta$ region, although strongly parameter dependent. Another sparticle decay mode could go through the lightest neutralinos, making the Higgs boson ``invisible''~\cite{Meisel}. The best performance for such a search mode is expected with the difficult VBF production. Instead of decays into sparticles, it is also possible to look for the Higgs bosons being produced in cascade decays from sparticles. Again, such searches are very parameter dependent.

\begin{figure}[t]
  \begin{center}
    \begin{minipage}[t]{.48\textwidth}
      \begin{center}
        \includegraphics[scale=0.42]{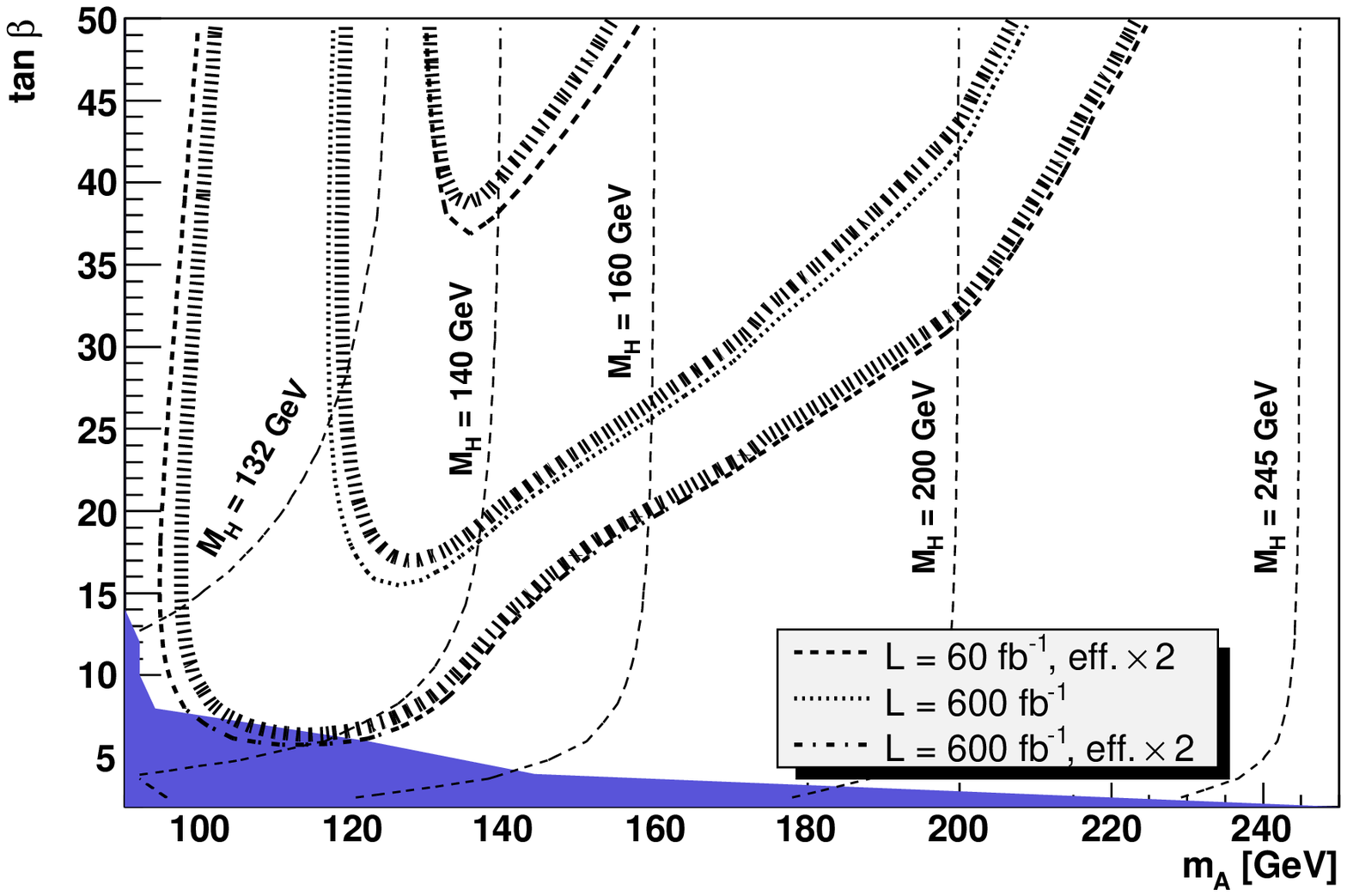}
        \vspace{-3mm}
        \caption{FP420 discovery reach for the MSSM ${\rm H}$ boson~\cite{Albrow:2008pn}.}
        \label{fig:fp420hbb}
      \end{center}
    \end{minipage}
    \hspace{\stretch{1}}
    \begin{minipage}[t]{.48\textwidth}
      \begin{center}
        \includegraphics[scale=0.46]{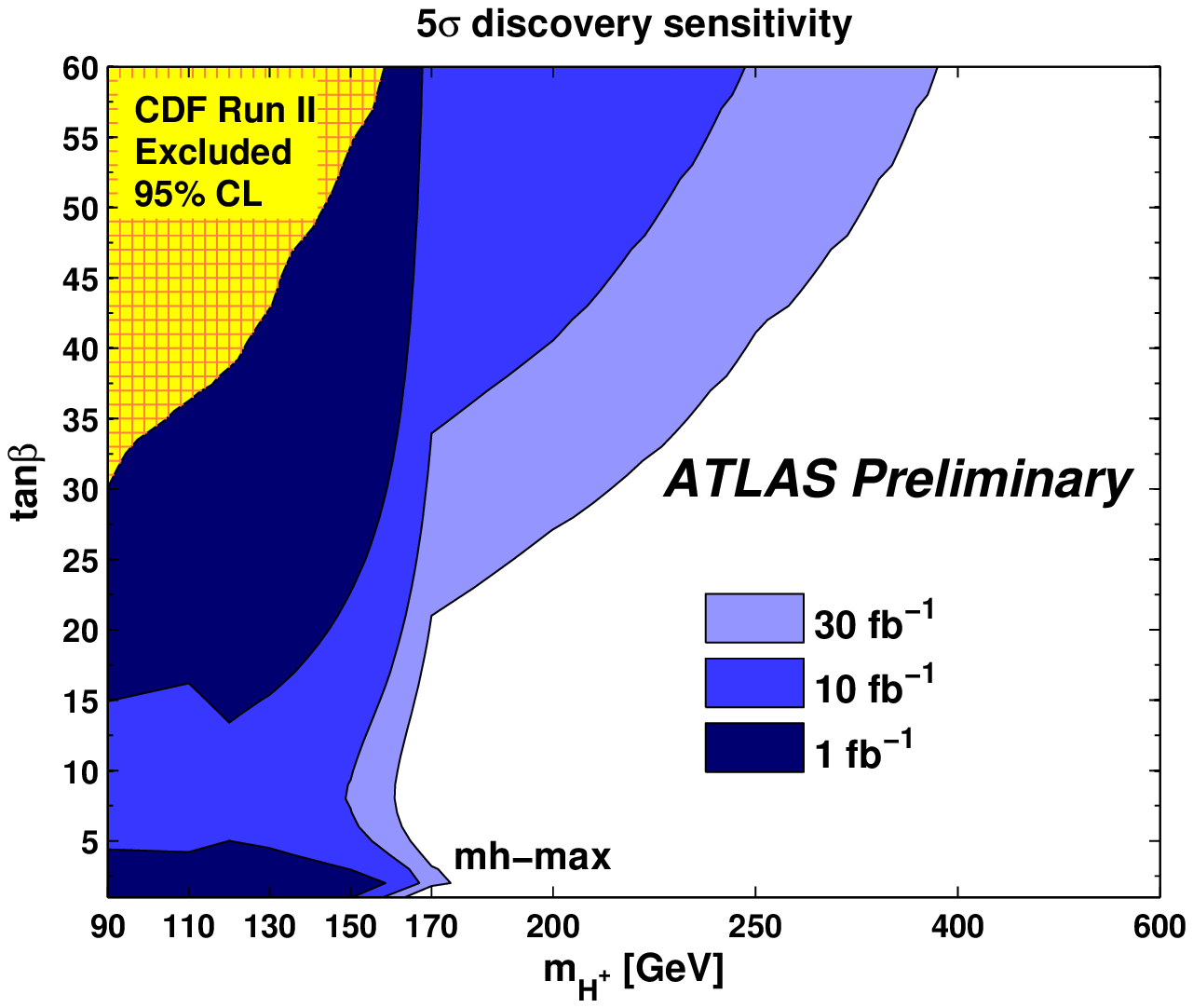}
        \vspace{-3mm}
        \caption{ATLAS sensitivity for the MSSM ${\rm H^{\pm}}$ boson~\cite{ATLASPerformance}.}
        \label{fig:atlaschhiggscontours}
      \end{center}
    \end{minipage}
  \end{center}
  \vspace{-5mm}
\end{figure}

The ${\rm h \to b\bar{b}}$ decay, almost hopeless inclusively or in ${\rm t\bar{t}}$ or ${\rm b\bar{b}}$ associated production, can become a viable channel in central exclusive production, where the cross section can be about $20 \, {\rm fb}^{-1}$. The two deflected protons are tagged in the above described FP420 detectors, and two central jets are detected in an otherwise very quiet environment. The main challenge for this search is the trigger, which cannot profit from the proton signal in the roman pots for the low-latency Level-1 decision. Also precise timing (order $10 \, {\rm ps}$ of the proton arrival is needed to resolve the primary vertex amidst pile-up interactions at high luminosity. An FP420-CMS combined study~\cite{Albrow:2008pn} shows a large discovery reach in the MSSM parameter space provided sufficient integrated luminosity is accumulated. For the heavier $CP$-even Higgs boson ${\rm H}$ the expected reach is shown in Figure~\ref{fig:fp420hbb}.

\subsubsection*{Charged MSSM Higgs-Boson Searches}

The production and decay of the charged Higgs boson at tree level is controlled by its mass and by $\tan\beta$. The main production mechanism ${\rm gg\to t H^{\pm} X}$ has a large cross section at both low and high $\tan\beta$, decreasing with increasing $m_{\rm H^{\pm}}$, and with a transition at around the top-quark mass. For high $m_{\rm H^{\pm}}$ the dominant decay is ${\rm H^{\pm} \to tb}$, which is very hard to exploit because of the huge ${\rm t\bar{t}}$+jets background~\cite{Lowette:2006bs}. The decay ${\rm H^{\pm} \to \tau\nu}$, subleading at large $\tan\beta$, is dominant below the top-quark mass, and therefore the golden channel for the entire mass range. The overall discovery reach is summarized in Figure~\ref{fig:atlaschhiggscontours}.

Below the top quark mass the search concentrates on identification of a $\tau$ jet, MET and ${\rm b}$ jets, with the trigger provided by the lepton from the ${\rm W}$~\cite{Baarmand:2006dm}. The main background comes from top and ${\rm W}$+jets production. Also a hadronic search is considered~\cite{Biscarat,ATLASPerformance}. Above the top-quark mass a fully hadronic strategy is followed, by both CMS and ATLAS~\cite{Mohn:2007fd,ATLASPerformance,Kinnunen}, which allows to reconstruct the ${\rm H^{\pm}}$ transverse mass and to fully constrain the top quark. The trigger is provided by a single energetic $\tau$ jet. The main backgrounds from top, ${\rm W}$+jets and QCD multijets are suppressed with an isolated muon and electron veto, a requirement for one ${\rm b}$-tagged jet and the presence of a large-$E_{\rm T}$ $\tau$ jet. Additionally, helicity correlations from the scalar versus vector nature of the mother particle can be used to suppress background $\tau$ jets, by requiring the leading track to carry more than $80\%$ of the $\tau$ energy.

\section{NON-MINIMAL HIGGS BOSON SEARCHES}

In this section a few of the searches for non-standard Higgs bosons beyond the Standard Model are described. The first case deals with $CP$ violation in the MSSM Higgs sector from complex trilinear couplings to squarks. As a consequence the usual $CP$ eigenstates mix into the physical states ${\rm H}_1$, ${\rm H}_2$ and ${\rm H}_3$. The so-called CPX scenario~\cite{Accomando:2006ga} has been studied in detail, in which the $CP$-violating effect is maximized. In this scenario the constraints from LEP on the low $\tan\beta$ region are much relaxed. At the LHC the relevant channels for the lightest ${\rm H_1}$ are the VBF production, with subsequent decays into ${\rm W^{+}W^{-}}$ and ${\rm \tau^{+}\tau^{-}}$, the ${\rm b}$-associated production with ${\rm H_1 \to \mu^{+}\mu^{-}}$ and the difficult $\rm t\bar{t} H_1$, $\rm H_1 \to b\bar{b}$. As in the regular MSSM at least one Higgs boson is observable in the full parameter space, except for a small uncovered region at low $m_{\rm H^{\pm}}$ and low $\tan\beta$.

Non-minimal models with a Higgs-boson triplet give rise to a doubly charged Higgs boson ${\rm \phi^{\pm\pm}}$. Examples are the littlest Higgs model and left-right symmetric models. ATLAS~\cite{Azuelos:2004dm} pursued a search in VBF production, with ${\rm \phi^{\pm\pm} \to W^\pm W^\pm \to \ell^\pm \ell^\pm}$, characterized by high-momentum same-sign leptons and forward jets. On the CMS side a search for pair production was performed with decays into $\mu^\pm \mu^\pm$, $\mu^\pm \tau^\pm$ and $\tau^\pm \tau^\pm$~\cite{Rommerskirchen}.

In Randall-Sundrum models with extra dimensions, finally, the scalar radion field $\phi$ can give rise to a sizeable ${\rm \phi \to hh}$ branching fraction in certain regions of parameter space. The golden search strategy ${\rm \phi \to hh \to \gamma\gamma bb }$ uses the di-photon trigger to collect the events~\cite{Dominici}. The background from $\gamma\gamma$+jets can be suppressed using ${\rm b}$-tagging, and both the Higgs boson mass peaks can be reconstructed. In the less clean ${\rm \phi \to hh \to \tau\tau bb}$ one isolated lepton, one $\tau$ jet and two ${\rm b}$ jets are identified. Backgrounds come in this case from ${\rm t\bar{t}}$, $\rm Z$+jets and $\rm W$+jets.

\section{CONCLUSIONS}

The LHC experiments are well equipped and ready to search for BSM Higgs-boson signals. If nature is like the MSSM, then at least one Higgs boson will be found. Many of the discovery channels require several years of data taking though, but some channels can possibly provide evidence or lead to large exclusions rapidly. Apart from the MSSM searches, the LHC experiments also prepare to look for unexpected, more exotic Higgs-boson signals.

\begin{acknowledgments}
This work was made possible by the support of the U.S. Department of Energy, Grant No. DEFG03-91ER40618.
\end{acknowledgments}

\end{document}